\documentstyle[aps,epsfig]{revtex}
\tighten
\begin{document}
\draft
\preprint{Preprint Numbers: \parbox[t]{45mm}{nucl-th/0102057}}

\title{$K_{l3}$ transition form factors}

\author{Chueng-Ryong Ji and Pieter Maris} 
\address{Department of Physics, North Carolina State University, 
        Raleigh,  NC 27695-8202}
\date{\today}
\maketitle
\begin{abstract}
The rainbow truncation of the quark Dyson--Schwinger equation is
combined with the ladder Bethe--Salpeter equation for the meson bound
state amplitudes and the dressed quark-W vertex in a manifestly
covariant calculation of the $K_{l3}$ transition form factors and decay
width in impulse approximation.  With model gluon parameters previously
fixed by the chiral condensate, the pion mass and decay constant, and
the kaon mass, our results for the $K_{l3}$ form factors and the kaon
semileptonic decay width are in good agreement with the experimental
data.
\end{abstract}

%

\section{Introduction}
The central unknown quantity required for reliable calculations of weak
decay amplitudes are the hadronic matrix elements.  In this respect, the
semileptonic $K_{l3}$ decay is an interesting process: pseudoscalar
mesons are well-understood as Goldstone bosons, only the vector part of
the weak interaction is involved, and there are experimental data
available which allows one to judge various theoretical (model)
calculations.  These calculations generally fall into two different
classes: effective theories using meson degrees of freedom, e.g. chiral
perturbation theory~\cite{Leutwyler:1984je,Gasser:1985ux}, and models
employing quark and gluon degrees of freedom, e.g. constituent quark
models~\cite{Isgur:1975hu,Choi:1999jd,Choi:1999nu} and the
Dyson--Schwinger
approach~\cite{Afanasev:1997my,Kalinovsky:1997ii,Ivanov:1999ms}.

In particular, the light-front approach has been a popular framework for
a Hamiltonian approach to analyze exclusive hadronic
processes~\cite{Brodsky:1998de}, and the pseudoscalar-to-pseudoscalar
weak transition form factors have been studied within a light-front
constituent quark model in Ref.~\cite{Choi:1999jd,Choi:1999nu}.  One
complication of the light-front formalism is that in the analysis of
timelike exclusive processes one needs to take into account light-front
nonvalence contributions, which are absent in a manifestly Poincar{\`e}
invariant approach.  Recently, an effective treatment has been
presented~\cite{Ji:2000fy} to incorporate such contributions, and a
systematic program has been laid out~\cite{Brodsky:1999hn} to take into
account the nonvalence contributions.  However, the systematic program
explicitly requires all higher Fock-state components while there has
been relatively little progress in computing the basic wave-functions of
hadrons from first principles. Furthermore, the (nonperturbative)
dressing of the quark-W vertex and the dynamical chiral symmetry
breaking are yet underdeveloped aspects in this approach. In particular
the latter aspect needs to be seriously considered for processes
involving pions and kaons, which are the pseudo-Goldstone bosons
associated with dynamical chiral symmetry
breaking~\cite{Maris:1998hd,Maris:1997tm}.

In this work we use the set of Dyson--Schwinger equations [DSEs] to
calculate the $K_{l3}$ transition form factors.  The DSEs provide us
with a manifestly covariant approach which is consistent with dynamical
chiral symmetry breaking~\cite{dcsb}, electromagnetic current
conservation~\cite{Maris:2000bh}, and quark and gluon
confinement~\cite{conf}.  For recent reviews on the DSE and its
application to hadron physics, see
Refs.~\cite{Roberts:2000aa,Alkofer:2000wg}.  Our calculation of the
$K_{l3}$ transition form factors is analogous to recent calculations of
the pion and kaon electromagnetic form factors in impulse
approximation~\cite{Maris:2000bh}.  Here we demonstrate that in the same
framework, with a consistent dressing of the quark-W vertex, we can also
describe the semileptonic decay of both the neutral and charged kaons,
without any readjustment of the model parameters.  In Sec.~II we review
the formulation that underlies a description of the meson form factors
within a modeling of QCD through the DSEs, and discuss the details of
the model.  In Sec.~III we present our numerical results for the form
factors and decay widths.  A discussion of our results is given in
Sec.~IV.

\section{Meson form factors within the DSE approach}
The matrix elements describing kaon semileptonic decays are
\begin{eqnarray}
 J_\mu^{K^+}(P,Q) &=& 
        \langle \pi^0(p) | \bar {s}\gamma_\mu u | K^+(k)\rangle
\nonumber \\ &=& \frac{1}{\sqrt{2}} 
        \left( f_+(-Q^2) P_\mu + f_-(-Q^2) Q_\mu \right) \,,
\end{eqnarray}
for the charged kaon, and for the neutral kaon
\begin{eqnarray}
 J_\mu^{K^0}(P,Q) &=& 
        \langle \pi^-(p) | \bar {s}\gamma_\mu u | K^0(k)\rangle
\nonumber \\ &=&
        f_+(-Q^2) P_\mu + f_-(-Q^2) Q_\mu \,,
\end{eqnarray}
where $P_\mu = (p+k)_\mu$ and $Q_\mu = (k-p)_\mu$, with $P\cdot Q = 
m_\pi^2-m_k^2$ for on-shell pions and kaons\footnote{We employ a
Euclidean space formulation with $Q^2 < 0$ for timelike vectors,
${\gamma_\mu,\gamma_\nu} = 2\delta_{\mu\nu}$, and $\gamma_\mu^\dagger =
\gamma_\mu$.}.  Alternatively, we can decompose $J_\mu^{K^0}$ into its
transverse and longitudinal components
\begin{eqnarray}
 J_\mu^{K^0}(P,Q) &=& f_+(-Q^2) \; T_{\mu\nu}(Q) \, P_\nu
                + f_0(-Q^2) \,\frac{P\cdot Q}{Q^2} \; Q_\mu \,,
\label{fzerodef}
\end{eqnarray}
with $T_{\mu\nu}$ the transverse projection operator
\begin{eqnarray}
 T_{\mu\nu}(Q) & = &
        \left(\delta_{\mu\nu} - \frac{Q_\mu Q_\nu}{Q^2} \right) \,.
\end{eqnarray}
In the isospin-symmetric limit, which we employ here, the form factors
$f_\pm(t=-Q^2)$ and $f_0(t)$ are the same for the $K^+$ and the $K^0$.

In impulse approximation, these matrix elements are given by
\begin{eqnarray}
J_\mu^{K^0}(P,Q) &=&
        N_c\int\!\!\frac{d^4q}{(2\pi)^4} \,
        {\rm Tr}\big[ S^d(q) \, \Gamma_{K}^{d\bar{s}}(q,q-k) 
        S^s(q-k) \, i \Gamma^{s\bar{u}W}_\mu(q-k,q-p)\, 
        S^u(q-p) \, \bar\Gamma_{\pi}^{u\bar{d}}(q-p,q) \big] \;, 
\label{impulse}
\end{eqnarray}
where $S^f$ is the dressed quark propagator with flavor index $f$,
$\Gamma_{\pi\,(K)}$ the pion (kaon) Bethe--Salpeter amplitude [BSA], and
$\Gamma^{s\bar{u}W}_\mu$ the dressed $s\bar{u}W$ vertex, each satisfying
their own DSE.  Note that the coupling constant and the CKM matrix
element $V_{us}$ are removed from the definition of the quark-W vertex.

\subsection{Dyson--Schwinger Equations}
The DSE for the renormalized quark propagator in Euclidean space is
\begin{eqnarray}
\label{gendse}
 S(p)^{-1} &=& i \, Z_2\, /\!\!\!p + Z_4\,m(\mu) 
        + Z_1 \int^\Lambda\!\!\!\frac{d^4q}{(2\pi)^4} \,g^2 
        D_{\mu\nu}(k) \frac{\lambda^a}{2}\gamma_\mu 
                S(q)\Gamma^a_\nu(q,p) \,,
\end{eqnarray}
where $D_{\mu\nu}(k)$ is the dressed-gluon propagator,
$\Gamma^a_\nu(q;p)$ the dressed-quark-gluon vertex, and $k=p-q$.  
The most general solution of Eq.~(\ref{gendse}) has the form
\mbox{$S(p)^{-1} = i /\!\!\! p A(p^2) + B(p^2)$} and is renormalized 
at spacelike $\mu^2$ according to \mbox{$A(\mu^2)=1$} and
\mbox{$B(\mu^2)=m(\mu)$} with $m(\mu)$ the current quark mass.
The notation $\int^{\Lambda}$ represents a translationally invariant
regularization of the integral with $\Lambda$ the regularization
mass-scale; at the end of all calculations this regularization scale can
be removed.

Mesons can be studied by solving the homogeneous Bethe--Salpeter
equation [BSE] for $q^a \bar{q}^b$ bound states, with $a$ and $b$ flavor
indices,
\begin{eqnarray}
 \Gamma^{a\bar{b}}_H(p_+,p_-) &=& 
        \int^\Lambda\!\!\!\frac{d^4q}{(2\pi)^4} \, K(p,q;Q) 
        \otimes S^a(q_+) \, \Gamma^{a\bar{b}}_H(q_+,q_-) \, S^b(q_-)\, ,
\label{homBSE}
\end{eqnarray}
where $p_+ = p + \eta Q$ and $p_- = p - (1-\eta) Q$ are the outgoing
and incoming quark momenta respectively, and similarly for $q_\pm$.
The kernel $K$ is the renormalized, amputated $q\bar q$ scattering
kernel that is irreducible with respect to a pair of $q\bar q$ lines.
This equation has solutions at discrete values of $Q^2 = -m_H^2$,
where $m_H$ is the meson mass.  Together with the canonical
normalization condition for $q\bar q$ bound states, it completely
determines $\Gamma_H$, the bound state BSA.  The different types of
mesons, such as (pseudo-)scalar, (axial-)vector, and tensor mesons,
are characterized by different Dirac structures.  The most general
decomposition for pseudoscalar bound states is~\cite{Maris:1997tm}
\begin{eqnarray}
\label{genpion}
\Gamma_{PS}(k_+,k_-) &=& \gamma_5 \big[ i E(k^2;k\cdot Q;\eta) 
        + \;/\!\!\!\! Q \, F(k^2;k\cdot Q;\eta) 
        + \,/\!\!\!k \, G(k^2;k\cdot Q;\eta) 
        + \sigma_{\mu\nu}\,Q_\mu k_\nu \,H(k^2;k\cdot Q;\eta) \big]\,,
\end{eqnarray}
where the invariant amplitudes $E$, $F$, $G$ and $H$ are Lorentz scalar
functions of $k^2$ and $k\cdot Q$.  Note that these amplitudes explicitly 
depend on the momentum partitioning parameter $\eta$.  However, 
so long as Poincar\'e invariance is respected, the resulting physical
observables are independent of this parameter~\cite{Maris:2000bh}.

In order to describe weak (and electromagnetic) form factors one also
needs the dressed $q\bar q W$ (and $q\bar q\gamma$) vertices.  These
vertices satisfy an inhomogeneous BSE: e.g. the $s\bar{u} W$ vertex
\mbox{$\Gamma^{u\bar{s}W}_\mu(p_+,p_-)$} satisfies
\begin{eqnarray}
 \Gamma^{s\bar{u}W}_\mu(p_+,p_-) &=& Z_2 \, 
                (\gamma_\mu-\gamma_\mu\gamma_5) 
        + \int^\Lambda\!\!\!\frac{d^4q}{(2\pi)^4} \, K(p,q;Q) 
        \otimes S^{s}(q_+) \, \Gamma^{s\bar{u}W}_\mu(q_+,q_-) 
                \, S^{s}(q_-)\, .
\label{verBSE}
\end{eqnarray}

\subsection{Model Truncation}
\label{modelcalc}
To solve the BSE, we use a ladder truncation, with an effective
quark-antiquark interaction that reduces to the perturbative running
coupling at large momenta~\cite{Maris:1997tm,Maris:1999nt}  
\begin{equation}
\label{ourBSEansatz}
        K(p,q;P) \to
        -{\cal G}(k^2)\, D_{\mu\nu}^{\rm free}(k)
        \textstyle{\frac{\lambda^i}{2}}\gamma_\mu \otimes
        \textstyle{\frac{\lambda^i}{2}}\gamma_\nu \,,
\end{equation}
where $D_{\mu\nu}^{\rm free}(k=p-q)$ is the free gluon propagator in
Landau gauge.  The corresponding rainbow truncation of the quark DSE,
Eq.~(\ref{gendse}), is given by \mbox{$\Gamma^i_\nu(q,p) \rightarrow 
\gamma_\nu\lambda^i/2$} together with \mbox{$Z_1 g^2 D_{\mu \nu}(k) 
\rightarrow {\cal G}(k^2) D_{\mu\nu}^{\rm free}(k) $}.
This combination of rainbow and ladder truncation preserves both the
vector and axial-vector Ward--Takahashi identities.  This ensures that
pions are (almost) massless Goldstone bosons associated with dynamical
chiral symmetry breaking~\cite{Maris:1997tm}; in combination with
impulse approximation for meson form factors, Eq.~(\ref{impulse}), it
also ensures current conservation.  Furthermore, this truncation was
found to be particularly suitable for the flavor octet pseudoscalar and
vector mesons since the next-order contributions in a quark-gluon
skeleton graph expansion, have a significant amount of cancellation
between repulsive and attractive corrections~\cite{Bender:1996bb}.

The model is completely specified once a form is chosen for the
``effective coupling'' ${\cal G}(k^2)$.  We employ the
Ansatz~\cite{Maris:1999nt}
\begin{eqnarray}
\label{gvk2}
\frac{{\cal G}(k^2)}{k^2} &=&
        \frac{4\pi^2\, D \,k^2}{\omega^6} \, {\rm e}^{-k^2/\omega^2}
        + \frac{ 4\pi^2\, \gamma_m \; {\cal F}(k^2)}
        {\textstyle{\frac{1}{2}} \ln\left[\tau + 
        \left(1 + k^2/\Lambda_{\rm QCD}^2\right)^2\right]} \;,
\end{eqnarray}
with \mbox{$\gamma_m=12/(33-2N_f)$} and
\mbox{${\cal F}(s)=(1 - \exp\frac{-s}{4 m_t^2})/s$}.  
The ultraviolet behavior is chosen to be that of the QCD running
coupling $\alpha(k^2)$; the ladder-rainbow truncation then generates the
correct perturbative QCD structure of the DSE-BSE system of equations.
The first term implements the strong infrared enhancement in the region
\mbox{$0 < k^2 < 1\,{\rm GeV}^2$} phenomenologically 
required~\cite{Hawes:1998cw} to produce a realistic value for the chiral
condensate.  We use \mbox{$m_t=0.5\,{\rm GeV}$},
\mbox{$\tau={\rm e}^2-1$}, \mbox{$N_f=4$}, \mbox{$\Lambda_{\rm QCD} =
0.234\,{\rm GeV}$}, and a renormalization scale \mbox{$\mu=19\,{\rm
GeV}$} which is well into the perturbative
domain~\cite{Maris:1997tm,Maris:1999nt}.  The remaining parameters,
\mbox{$\omega = 0.4\,{\rm GeV}$} and \mbox{$D=0.93\,{\rm GeV}^2$} 
along with the quark masses, are fitted to give a good description of
the chiral condensate, $m_{\pi/K}$ and $f_{\pi}$.

Within this model, the quark propagator reduces to the perturbative
propagator in the ultraviolet region.  However, in the infrared region
both the wave function renormalization $Z(p^2) = 1/A(p^2)$ and the
dynamical mass function $M(p^2)=B(p^2)/A(p^2)$ deviate significantly
from the perturbative behavior, due to chiral symmetry breaking.  It is
interesting to note that the typical results obtained using the rainbow
DSE, a significant enhancement of $M(p^2)$ below $1\;{\rm GeV}^2$ and
also an enhancement of $A(p^2)$, have recently been confirmed by lattice
simulations~\cite{Skullerud:2000un}; with the present ladder DSE model,
the functions $Z$ and $M$ are in semiquantitative agreement with the
forms obtained in lattice simulations~\cite{Maris:2000zf}.

The vector meson masses and electroweak decay constants one obtains in
this model are in good agreement with experiments~\cite{Maris:1999nt}.
Without any readjustment of the parameters, one gets remarkable
agreement with the most recent Jlab data for
$F_\pi$~\cite{Volmer:2000ek}, and also for other electromagnetic charge
radii and form factors~\cite{Maris:2000bh,Maris:2000wz}.  The strong
decays of the vector mesons into a pair of pseudoscalar mesons are also
well-described within this model~\cite{jarecke}.  Here we apply the same
approach to the kaon semileptonic decay.

\section{Numerical results}
\begin{figure}[h]
\epsfig{figure=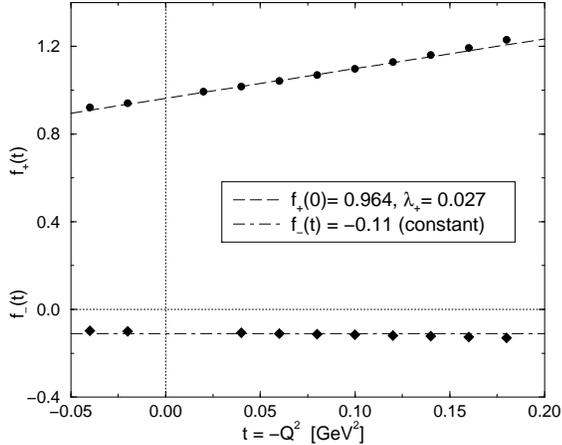,height=6cm}
\caption{Our results for $f_+(t)$ (solid dots) with linear fit 
(dashed line), and $f_-(t)$ (solid diamonds) with a constant fit 
(dot-dashed line) in the physical region.}
\label{fig1}
\end{figure}

To calculate the form factors $f_\pm(t)$, we start by solving the quark
DSE in rainbow approximation, and subsequently use its solution to solve
the homogeneous BSE, Eq.~(\ref{homBSE}), for the pseudoscalar bound
states, and the inhomogeneous BSE, Eq.~(\ref{verBSE}), for the quark-W
vertex, using the ladder truncation, Eq.~(\ref{ourBSEansatz}).  With
these BS amplitudes and quark propagators, we then calculate the form
factors in impulse approximation using Eq.~(\ref{impulse}).  In the
SU(3) flavor limit, $f_+ = F_\pi$, the pion electromagnetic form factor,
and $f_- = 0$.  We therefore expect $f_+$ to be of order one for small
$t$.  On the other hand, $f_-$, which is a measure of the constituent
mass ratio $M_s/M_u$~\cite{Isgur:1975hu}, is expected to be
significantly smaller.  This expectation is indeed confirmed by our
calculations, see Fig.~\ref{fig1}.

In the physical region, $m_l^2 < t < (m_K-m_\pi)^2 \simeq 0.13\;{\rm
GeV}^2$, our results for $f_{\pm}(t)$ are almost linear.  The
experimental data are often analyzed in terms of $f_+$ and $f_0$,
using a linear parametrisation
\begin{eqnarray}
       f(t) &=& f(0) \Big(1 + \frac{\lambda}{m_\pi^2}\; t\, \Big) \,.
\end{eqnarray}
Such a linear parametrisation for both $f_+$ and $f_0$ implies
that $f_-$ is independent of $t$, since $f_0$ is related to
$f_\pm$ via
\begin{eqnarray}
        f_0(t) &=& f_+(t) + \frac{t}{m_K^2-m_\pi^2} f_-(t) \,.
\end{eqnarray}
To within a few percent, $f_-$ is indeed almost constant in the physical
region, $0<t<0.13~{\rm GeV}^2$, as can be seen from Fig.~\ref{fig1}.
Our results are summarized in Table~\ref{table}, where we have included
a slope parameter $\lambda$ for $f_-(t)$ as well.

\begin{figure}[h]
\epsfig{figure=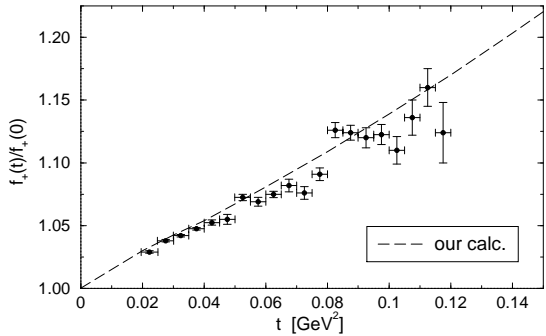,height=4.5cm}
\caption{Our results for $f_+(t)$, normalized to $f_+(0)=1$, 
compared to the experimental results~\protect\cite{Apostolakis:2000gs}.}
\label{fig2}
\end{figure}

Our results agree quite well with the available data for $f_+(t)$, see
Fig.~\ref{fig2}.  Also the results for other observables compare
reasonably well with the experimental data, given the error bars, and
the differences between the form factor parameters extracted from the
charged and the neutral kaon decay, see Table~\ref{table}.  The partial
decay width for $K\rightarrow \pi\,e\nu_e$ and
$K\rightarrow\pi\,\mu\nu_\mu$ is obtained by integrating the decay rate
\begin{eqnarray}
 \frac{d\Gamma}{dt} &=& \frac{G_F\,|V_{us}|^2}{24\pi^3}
 \left(1-\frac{m_l^2}{t}\right)^2 K(t)
        \Bigg[K^2(t) \, \left(1+\frac{m_l^2}{2\,t} \right) \; |f_+(t)|^2
        + m_K^2 \, \left(1-\frac{m_\pi^2}{m_K^2}\right)^2 \,
                \frac{3\,m_l^2}{8\,t} \; |f_0(t)|^2 \Bigg] \;,
\end{eqnarray}
where
\begin{eqnarray}
 K(t) &=& \frac{1}{2 m_K} \left((m_K^2+m_\pi^2-t)^2 
                - 4 m_K^2 m_\pi^2\right)^{\frac{1}{2}} \,,
\end{eqnarray}
and $m_l$ is the lepton mass.  Again, we find good agreement with the
data, see Table~\ref{table}.

For comparison, we have also included the results from some other model
calculations~\cite{Kalinovsky:1997ii,Ji:2000fy} and from 1-loop chiral
perturbation theory~\cite{Gasser:1985ux,Bijnens:1994me}.  Our results
compare quite well with chiral perturbation theory, not only for the
slope parameters, but also for the curvature of the form factors.  
A prediction from current algebra is the value of $f_0(t)$ at the
Callan--Treiman point, $t = m^2_K - m^2_\pi$, namely $f_K/f_\pi =
1.22~$\cite{Gasser:1985ux,callantreiman}.  The reason we get $f_0(m^2_K
- m^2_\pi) = 1.18$ instead of $1.22$ is related to our value of $f_K$:
we obtain $f_K = 109~{\rm MeV}$ in our
model~\cite{Maris:1997tm,Maris:1999nt}, compared to the experimental
value $113~{\rm MeV}$.  Because our approach satisfies constraints
coming from current conservation and the axial-vector Ward--Takahashi
identity, it is not surprising that we indeed agree with these
predictions from chiral perturbation theory.

\begin{figure}[h]
\epsfig{figure=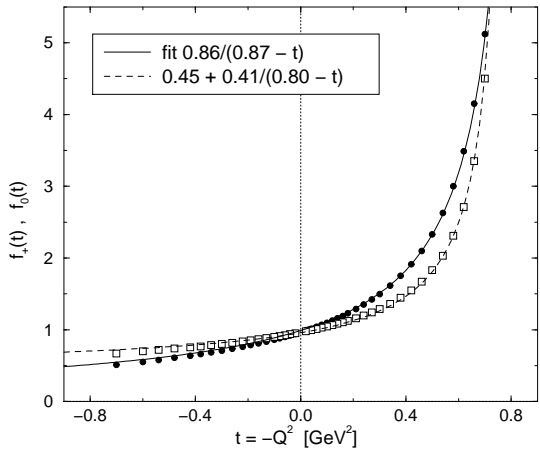,height=6cm}
\caption{Our results for $f_+(t)$ (solid dots) with monopole fit 
(solid curve), and $f_0(t)$ (open squares) with a constant plus 
monopole fit(dashed curve).}
\label{fig3}
\end{figure} 
Outside the physical region, $f_{\pm}(t)$ and $f_0(t)$ are clearly not
linear, as shown in Fig.~\ref{fig3}.  In the region we considered, our
result for $f_+(t)$ is very well approximated by a monopole, with
monopole mass $m^2 = 0.87\;{\rm GeV}^2$.  The reason for this monopole
behavior is easy to understand: the vertex function
$\Gamma^{s\bar{u}W}_\mu$ has a resonance peak in its transverse vector
components at the $K^*$ mass, because the homogeneous version of
Eq.~(\ref{verBSE}) has a solution at $Q^2 = -m_{K^*}^2$, corresponding
to the $K^*$ vector meson.  The longitudinal part of the quark-W vertex
has a resonance peak at the location of the $u\bar{s}$ scalar bound
state.  We tentatively identify this with the $K_0^*(1430)$ meson;
however, in contrast to vector states, the use of ladder approximation
is expected to be inadequate for scalars~\cite{beyond}.  Near these
bound states, the vertex behaves like
\begin{eqnarray}
 \Gamma^{s\bar{u}W}_\mu(p_+,p_-;Q) &\sim&
        \Gamma_\mu^{\hbox{\scriptsize regular}}(p_+,p_-;Q)      
  + \frac{r_V \; T_{\mu \nu}(Q)}{Q^2 + m_{K^*}^2} 
                \Gamma^{K^*}_\nu(p_+,p_-;Q)
  + \frac{r_S \; Q_\mu}{Q^2 + m_{K_0^*}^2} \Gamma^{K_0^*}(p_+,p_-;Q) \,,
\end{eqnarray}
where $\Gamma^{K^*}$ and $\Gamma^{K_0^*}$ are the bound state BSAs, and
$r_V$ and $r_S$ the residues of the vertex at these
poles~\cite{Maris:1998hd}.  In our model we have actual poles in the
quark-W vertex, rather than resonance peaks, because we have not
included meson loop corrections in our kernel, and therefore we do not
generate a width for the $K^*$ nor the $K^*_0$.  In the spacelike region
and at small timelike $t=-Q^2$ we expect these corrections to be small.
However, they will modify our results significantly close to the
resonance peak.

As can be seen from Eq.~(\ref{impulse}), these poles in
$\Gamma^{s\bar{u}W}_\mu$ lead a similar behavior of $J^{K^0}_\mu$ near
these poles.  From Eq.~(\ref{fzerodef}) it is clear that the vector
meson pole leads to a pole in $f_+(t)$, but not in $f_0(t)$, whereas the
scalar meson induces a pole in $f_0(t)$ but not in $f_+(t)$
\begin{eqnarray}
 f_0(-Q^2) & = &  
        \frac{ - Q_\mu \; J^{K^0}_\mu }{ m_K^2 - m_\pi^2 } \;,
\label{fzero}
\\
 f_+(-Q^2) & = &  
 \frac{\left(Q^2 P_\mu + (m_K^2 - m_\pi^2)\,Q_\mu \right) \, J^{K^0}_\mu}
        { Q^2 \, P^2 - (m_K^2 - m_\pi^2)^2} \;.
\label{fplus}
\end{eqnarray}
Note that $f_-(t)$ will in general be sensitive to both the scalar and
the vector meson bound state.

The present model has a vector bound state at $t= 0.876\;{\rm GeV}^2$
(cf. the experimental mass $m_{K^*}^2 = 0.796\;{\rm
GeV}^2$)~\cite{Maris:1999nt}.  Thus the monopole behavior of the form
factor $f_+(t)$ in Fig.~\ref{fig3} simply reflects the existence of this
pole.  A similar behavior was found in the pion and kaon electromagnetic
form factors~\cite{Maris:2000bh}.  The physical $f_+(t)$ can be expected
to rise more slowly with $t$ than our present calculations due to meson
loop corrections, and will develop a nonzero imaginary part above the
threshold for intermediate $\pi\,K$ states.

The scalar form factor $f_0(t)$ does also exhibit the expected pole
behavior, and we are able to identify the scalar and the vector poles
separately, see Fig.~\ref{fig3}.  The lowest $u\bar{s}$ scalar bound
state in this model has a mass of $m^2 = 0.80\;{\rm GeV}^2$, which is
rather low compared to the $K^*_0(1430)$ with a mass $m^2 = 2.04\;{\rm
GeV}^2$; on the other hand, it could be an indication that there exists
a light scalar resonance $\kappa(900)$, as has been
speculated~\cite{kappa}.  Something similar happens in in the $u/d$
quark sector: in rainbow-ladder truncation, one typically finds a scalar
$u/d$ bound state around $m = 0.6$ to $0.7\;{\rm
GeV}$~\cite{Jain:1993qh,Burden:1997nh,Maris:2001ig}, just below the
$\rho$ mass.  This bound state could either correspond to a broad
$\sigma$, or to the $f_0$ and/or $a_0$, in which case the calculated
mass is about 30 to 40\% too low.  It is known that the leading
perturbative corrections to the ladder kernel cancel to a large extent
in both the pseudoscalar and vector channel, but not in the scalar
channel; we therefore expect in the scalar channel significant repulsive
corrections to the ladder kernel which could increase the mass of the
scalar bound state~\cite{beyond}.  We also expect meson loop corrections
to be more important in the scalar channel than in the vector channel.
Therefore, our calculation for $f_0(t)$ should not be trusted
quantitatively in the timelike region beyond the Callan--Treiman point,
$t = 0.23~{\rm GeV}^2$.

In the spacelike asymptotic region $f_+$ and $f_0$ seem to behave
differently: a monopole fit does not work for $f_0$.  In order to fit
$f_0$, we either need to add a nonzero constant, or use at least two
monopoles.  Simple power counting indicates that $J_\mu^{K^0}$ scales
like $1/Q$; inserting this behavior in Eqs.~(\ref{fzero}) and
(\ref{fplus}) implies that $f_+(t)$ falls like $1/t$, but $f_0(t)$ goes
to a constant (up to logarithmic corrections) at large spacelike $t$
(remember that $P$ scales like $Q$ in the asymptotic region).  This
behavior of $f_+$ agrees with the pQCD predictions for the
electromagnetic pion and kaon form factor~\cite{Lepage:1979zb}; the
behavior of $f_0$ can be understood if one realizes that the combination
$f_0(t)\,\frac{m^2_K-m^2_\pi}{t}$ does fall like $1/t$ if $f_0(t)$ goes
to a constant.

\section{Discussion}
Using the rainbow-ladder truncation of the set of DSEs with a model for
the effective quark-antiquark interaction that has been fitted to the
chiral condensate and $f_\pi$, we study the $K_{l3}$ decay in impulse
approximation.  Our results, both for the form factors and for the decay
width, are in good agreement with experimental data, without any
readjustment of the parameters; they also compare quite well with chiral
perturbation theory.  Note however that in chiral perturbation theory
the pion charge radius is used as input, in order to fix a low-energy
constant, which is important for these form factors as well, in
particular for $\lambda_+$.  In our calculation, the only model
parameters are in the infrared behavior of the effective quark-antiquark
interaction, which were fitted to the chiral condensate and the pion
decay constant~\cite{Maris:1999nt}; the pion charge radius follows from
a calculation similar to the one presented here for the $K_{l3}$
decay~\cite{Maris:2000bh}.

Our approach is based on previous work by Kalinovsky {\it et
al.}~\cite{Kalinovsky:1997ii}, and the results are quite similar.
However, an important difference with this earlier study is that we
dress the quark-W vertex by solving the inhomogeneous BSE for this
vertex.  The main advantage of doing so is that we thus automatically
include effects coming from intermediate vector and scalar mesons.
Another difference is that here we use actual solutions of a DSE for the
propagators and BS amplitudes in Eq.~(\ref{impulse}), rather than
phenomenological parametrisations, which reduces the number of
parameters in the calculation.

We have demonstrated explicitly that both $f_+(t)$ and $f_0(t)$ exhibit
resonance peaks in the timelike region due to the existence of vector
and scalar bound states respectively.  The effects of meson loops are
not included in our present calculation, which is why we find a pole
behavior rather than a resonance peak.  In Ref.~\cite{Kalinovsky:1997ii}
it was already demonstrated that non-analytic effects from such loops
could contribute significantly to the behavior of the form factors, not
only beyond the threshold for $\pi K$ production, but also in the
physical region.  We hope to be able to incorporate these effects in
future work.

Our results are also similar to those obtained recently by Ji and
Choi~\cite{Ji:2000fy} in a light-front calculation, at least for
$f_+(t)$, which dominates this decay, even though details of the
calculation are quite different.  In light-front calculations of
timelike processes one has to include particle-number-nonconserving Fock
states to recover the Lorentz covariance, which complicates the
calculation~\cite{Ji:2000fy,Brodsky:1999hn}.  Since our approach is
manifestly covariant, such contributions are automatically included in
our impulse approximation.  Another difference is that we use
momentum-dependent quark self-energies, consistent with dynamical chiral
symmetry breaking, whereas in Ref.~\cite{Ji:2000fy} constituent quarks
with fixed masses are used.  Recently, the effects of a running quark
mass (instead of a fixed constituent mass) have been explored in a
light-front calculation of the pion form
factor~\cite{Kisslinger:2001gw}.  It would be interesting to see the
effect of such a running mass on the $K_{l3}$ form factors, in
particular on $f_-(t)$, which in general appears to be quite sensitive
to details of the calculation.

\section*{Acknowledgments}
We would like to thank P.C.~Tandy, C.D.~Roberts, K.~Maltman,
S.R.~Cotanch, and H.~Choi for stimulating discussions and useful
suggestions.  This work was supported by the US DOE under grants
No. DE-FG02-96ER40947 and DE-FG02-97ER41048, and benefited from the
resources of the National Energy Research Scientific Computing Center.


\begin{table}
\begin{center}
\begin{tabular}{l|c|cc|ccc}
                & our   & \multicolumn{2}{c|}{experiment}
                                & \multicolumn{3}{c}{other theory} \\
        
                & calc  & $K^+$ & $K^0$ 
                & $\chi$PT~\protect\cite{Gasser:1985ux,Bijnens:1994me}
                & DSE model~\protect\cite{Kalinovsky:1997ii} 
                & light cone~\protect\cite{Ji:2000fy} \\ \hline
$f_+(0)   $     & 0.964 &       &        
                                & 0.977    & 0.98  & 0.962 \\
$\lambda_+(e3)$ & 0.027 & $.0276\pm.0021$& $.0288\pm.0015$ 
                                & 0.031    & 0.028 & 0.026 \\
$\lambda_+(\mu3)$&0.027 & $.031 \pm.008$ & $.034 \pm.005$  
                                & 0.031    & 0.029 & 0.026 \\
$-f_-(0)  $     & 0.10  &   &   & 0.16     & 0.24  &       \\
$\lambda_-$     & 0.03  &   &   &          & 0.023 &       \\
$-\xi=-f_-(0)/f_+(0)$&0.11& $0.31\pm.15$ & $ 0.11\pm.09$  
                                & 0.17     & 0.25  & 0.01  \\
$\lambda_0$     & 0.018 &  $.006\pm.007$ & $ .025\pm.006$ 
                                & 0.017    & 0.007 & 0.025 \\
$f_0(m^2_K-m^2_\pi)$ & 1.18 &   &   & 1.22 & 1.18  &       \\ \hline
$\Gamma(K_{e3})$& 7.38  & 3.89 & 7.50  & & & 7.3   \\
$\Gamma(K_{\mu3})$&4.90 & 2.57 & 5.26  & & & 4.92  
\end{tabular}
\end{center}
\caption{Our results, compared to data~\protect\cite{PDG} where 
available and some other calculations.  The partial decay width is in 
$10^6\;{\rm s}^{-1}$; because $J_\mu^{K^+} = J_\mu^{K^0} / \sqrt{2}$
in the exact isospin limit, the experimental value for the $K^+$ partial
decay width should be multiplied by a factor of two, in order to compare
it with our calculation and with the $K^0$ partial decay width.}
\label{table}
\end{table}
%
\end{document}